\documentstyle[aps,prd,multicol,epsf,epsfig]{revtex}\begin{document}

\newcommand{\be}{\begin{equation}}
\newcommand{\ee}{\end{equation}}
\newcommand{\ba}{\begin{eqnarray}}
\newcommand{\ea}{\end{eqnarray}}
\title{Ultra-High Energy Cosmic Rays from Neutrino Emitting Acceleration Sources?}
\author{Oleg~E.~Kalashev$^a$, Vadim~A.~Kuzmin$^a$, Dmitry~V.~Semikoz$^b$,
G{\"u}nter Sigl$^c$}
\address{$^a$ Institute for Nuclear Research of the Academy
of Sciences of Russia,\\
Moscow, 117312, Russia\\
$^b$ Max-Planck-Institut f\"ur Physik (Werner-Heisenberg-Institut),\\
F\"ohringer Ring 6, 80805 M\"unchen, Germany\\
$^c$ Institut d'Astrophysique de Paris, C.N.R.S., 98 bis boulevard
Arago, F-75014 Paris, France}

\maketitle

\vspace{0.5truecm}
\begin{abstract}
We demonstrate by numerical flux calculations that neutrino beams producing the observed highest energy
cosmic rays by weak interactions with the relic neutrino background
require a non-uniform distribution of sources. Such sources
have to accelerate protons  at least up to $10^{23}$ eV, have to be
opaque to their primary protons, and should emit the secondary
photons unavoidably produced together with the neutrinos only
in the sub-MeV region to avoid conflict with the diffuse $\gamma-$ray
background measured by the EGRET experiment. Even if such a
source class exists, the resulting large
uncertainties in the parameters involved in this 
scenario does currently not allow to extract any meaningful
information on absolute neutrino masses.
\end{abstract}
\vspace{1truecm}

\begin{multicols}{2}

\narrowtext

\section{Introduction}
In acceleration scenarios ultra high energy cosmic rays (UHECRs) with
energies above $10^{18}\,$eV are assumed to be protons accelerated in
powerful astrophysical sources.  During their propagation, for
energies above $\gtrsim50$ EeV ($1EeV = 10^{18} eV$) they lose energy
by pion production and pair production (protons only) on the cosmic
microwave background (CMB). For sources further away than a few dozen
Mpc this would predict a break in the cosmic ray flux known as
Greisen-Zatsepin-Kuzmin (GZK) cutoff~\cite{gzk}, around
$50\,$EeV. This break has not been observed by experiments such as
Fly's Eye~\cite{Fly}, Haverah Park~\cite{Haverah},
Yakutsk~\cite{Yakutsk} and AGASA~\cite{AGASA}, which
instead show an extension beyond the expected GZK cutoff and events
above $100\,$EeV. However the new experiment HiRes~\cite{Hires}
currently seems to see a cutoff in the monocular data~\cite{Hires2001}.
Taking into account that all old experiments except perhaps AGASA do not
have sufficient statistics in the highest energy region to settle
the question, the existence of a possible cutoff remains unclear
at the moment. The apparent absence of a cutoff especially in the
AGASA data has in recent years triggered many
theoretical explanations ranging from conventional acceleration in
astrophysical sources to models invoking new physics such as the
top-down scenarios in which energetic particles are produced in the
decay of massive relics from the early Universe~\cite{bs}. This enigma
has also fostered the development of large new detectors of ultra-high energy
cosmic rays which will increase very significantly the statistics at
the highest energies~\cite{reviews}.

In bottom-up scenarios of UHECR origin, in which protons are
accelerated in powerful astrophysical objects such as hot spots of
radio galaxies and active galactic nuclei~\cite{biermann}, one would
expect to see the source in the direction of arrival of UHECRs,
but above the GZK cutoff in general no suitable candidates have been
found within the typical energy loss distance of a few tens of Mpc for
the known electromagnetically or strongly interacting
particles~\cite{ssb,ES95}. Even assuming significant deflection
by large scale extragalactic magnetic fields requires at
least several sources~\cite{ils} whose locations have not been
identified yet. 

Moreover, recent observations of small scale clustering by the AGASA
experiment~\cite{AG1} suggest that sources of UHECR are
point-like~\cite{anis1,anis2}. 
This fact together with the lack of nearby sources favors the possibility of
sources much further away than 100 Mpc, at redshifts of order
unity. An additional motivation for
this possibility comes from recently reported possible correlations
of the arrival directions of observed UHECR above $\simeq50\,$EeV
with certain classes of sources such as compact radio
galaxies~\cite{corr_radio} or BL Lacertae objects~\cite{corr_bllac}. 
In the latter case it is still possible that the sources are located at 
moderate distances $z\simeq0.1$. In this case photons with
extremely high energies $E>10^{23}$ eV  can  propagate
several hundred Mpc (constantly loosing energy) 
and can create secondary photons inside the GZK volume~\cite{photons}. 
However, this model requires both extreme energies of primary photons
and extremely small extra galactic magnetic fields (EGMFs)
$B\lesssim10^{-12}$~G. Moreover, if a correlation with any source at
redshift $z > 0.2$ is found, this model will be ruled out.  
 
If sources of the highest energy cosmic rays are
indeed at cosmological distances $z \sim 1$, the only known mechanism
not involving new physics except for neutrino masses assumes
neutrinos as messenger particles: Charged particles accelerated
in such sources give rise to a secondary neutrino beam which can propagate
essentially unattenuated. If this neutrino beam is sufficiently strong
it can produce the observed UHECRs within 100 Mpc by electroweak (EW)
interactions with the relic neutrino background~\cite{zburst1}.
Specifically, if the relic neutrinos have a mass $m_\nu$, Z-bosons,
whose decay products can contribute to the UHECR flux, can be
resonantly produced by ultra high energy (UHE) neutrinos of energy
$E_\nu\simeq M_Z^2/(2m_\nu)\simeq4.2\times10^{21}\,{\rm eV}\,({\rm eV}/m_\nu)$.

However, this ``Z-burst'' mechanism is severely constrained by at least
two types of observational data: First, there are upper limits on the
UHE neutrino flux, based on the non-observation of horizontal
air showers by the old Fly's Eye experiment~\cite{baltrusaitis} 
or by the AGASA experiment \cite{AgAsA}
and from the non-observation of radio pulses that would
be emitted from the showers initiated by the UHE neutrinos
on the moons rim~\cite{goldstone}. Second, even if the sources exclusively
emit neutrinos, the EW interactions also produce photons and
electrons which initiate an electromagnetic (EM) cascade which transfers
the injected energy down to below the pair production threshold
for photons on the CMB~\cite{bs}. The cascade thus gives rise to a
diffuse photon flux in the GeV range which is constrained by the flux observed
by the EGRET instrument on board the Compton $\gamma-$ray
observatory~\cite{egret}. Reproducing the observed UHECR
flux by the Z-burst mechanism under these 
two constraints has been shown to in general require local relic neutrino 
over-densities in order to increase the local UHECR flux. These
over-densities turn out to be much higher than values 2--3 which
would be expected from the over-density in the local
supercluster~\cite{zburst2}.

In order to avoid this difficulty one can suppose that the Z-burst
mechanism is responsible only for part of the UHECR flux~\cite{fkr}.
In this case, one can reduce both primary neutrino and secondary photon
fluxes and obey all existing limits. However, the price for this
is to explain only a part of the UHECR events by the Z-burst mechanism
and the necessity for a second source mechanism for UHECRs.

Furthermore, Ref.~\cite{fkr} claims that already
the present data provides possible evidence for the relic
neutrino background and starts to constrain the
absolute neutrino mass, a possibility that has recently been
discussed in principle in Ref.~\cite{pw}. This claim is based
on tuning many unknown parameters such as the value of the EGMF,
the universal radio background (URB) which governs pair production
of UHE $\gamma-$rays, and the
neutrino source distribution. Also, Ref.~\cite{fkr}
did not take into account propagation of UHE photons, instead
assuming that all photons are down-scattered into the GeV region. 
In addition, simply due to the much larger
statistics at lower energies, the quality of the fits
performed in Ref.~\cite{fkr} is dominated by the low-energy background
component. Finally,  Ref.~\cite{fkr} assumed that the sources
do not emit any $\gamma-$rays, although the $\gamma-$ray energy
fluence produced by pion production of accelerated nuclei should be
comparable to the produced neutrino fluence, as will be
discussed in Sec.~\ref{phot+nu}. 

In the present paper we show that for all neutrino
masses in the range $ 0.07 {\rm eV}\leq m_\nu\leq 1 {\rm eV}$ 
one can find parameters that fit the UHECR observations
with comparable quality. We therefore conclude that, at least
at the current state of knowledge, it is impossible to extract
evidence for the relic neutrino background or even best fit values
for absolute neutrino masses from UHECR data.

We do not consider in the present paper neutrino interaction channels 
with multiple $W^{\pm}$ and/or $Z^0$ production. These channels could be important in
case of  neutrino masses $m \gtrsim 3$ eV \cite{Fargion}, which however are
strongly disfavored by considerations on large scale structure formation \cite{Fukugita}.

By detailed numerical flux calculations we show that a non-uniform  
source distribution allows the  Z-burst mechanism to explain
the UHECR flux without substantial relic neutrino over-densities.
However, this only works if the sources exclusively emit
neutrinos. Because isospin symmetry requires the energy fluence
of neutrinos and $\gamma-$rays produced by hadronic charged
primary interactions in the source are comparable, this will
require the photons to be down-scattered below the GeV range
within the source.

\section{Neutrino source}
We assume in this section that a pure neutrino source model
can somehow be constructed and start with this case.
In the next section we relax this condition and include other
primary particles into consideration.

Our simulations are based on two independent codes that
have extensively been compared down to the level of
individual interactions. Both of them are implicit transport codes
that evolve the spectra of nucleons, $\gamma-$rays, electrons,
electron-, muon-, and tau-neutrinos, and their antiparticles
along straight lines. Arbitrary injection spectra and
redshift distributions can be specified for the sources and
all relevant strong, electromagnetic, and weak interactions
have been implemented. For details see Refs.~\cite{code1,code2}.
Specifically relevant for neutrino interactions in the current
problem are both the s-channel production of
Z bosons and the t-channel production of W bosons.
The decay products of the Z boson were taken from simulations
with the ~\cite{PYTHIA} Monte Carlo event generator using the
tuned parameter set of the OPAL Collaboration~\cite{OPAL}.
The main ambiguities in propagation concern the unknown
rms magnetic field strength $B$ which can influence the
predicted $\gamma-$ray spectra via synchrotron cooling
of the electrons in the EM cascade, and the strength of the
URB which influences pair production
by UHE $\gamma-$rays~\cite{pb}. Photon interactions in the GeV to
TeV range are dominated by infrared and optical universal photon 
backgrounds (IR/O), for which we took the results of Ref. \cite{Primack}. 
The resulting photon flux in GeV range in not sensitive to
details of the IR/O backgrounds.

Predictions for the nucleon
fluxes agree within tens of percents whereas photon fluxes
agree only within a factor $\simeq2$ between the two codes.
The latter mostly reflects the ambiguities in photon propagation
mentioned above, but has no influence on the conclusions
of this paper.

For the present investigation we parameterize the neutrino
injection spectra per comoving volume in the following way:
\begin{eqnarray}
  \phi_\nu(E,z)&=&f(1+z)^m\,E^{-q_\nu}\Theta(E^\nu_{\rm max}-E)\,
  \nonumber\\
  &&z_{\rm min}\leq z\leq z_{\rm max}\,,\label{para_nu}
\end{eqnarray}
where $f$ is the normalization that has to be fitted to the
data. The free parameters are the spectral index $q_\nu$, the maximal
neutrino energy $E^\nu_{\rm max}$, the minimal and maximal
redshifts $z_{\rm min}$, $z_{\rm max}$, and the redshift
evolution index $m$. We assume for simplicity that all six
neutrino species (three flavors including antiparticles) are
completely mixed as suggested by experiments~\cite{superK}
and thus have equal fluxes given by Eq.~(\ref{para_nu}).
Finally we chose the Hubble parameter
$H_0 = 70~ {\rm km}~{\rm s}^{-1}~ {\rm Mpc}^{-1}$ and a
cosmological constant $\Omega_\Lambda = 0.7$, as favored today.

\begin{figure}[ht]
\centering\leavevmode \epsfxsize=3.5in \epsfbox{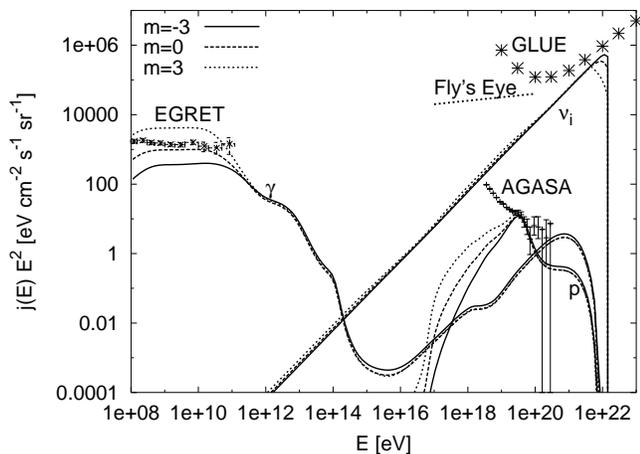}
\caption[...]{
Fluxes of neutrinos, $\gamma-$rays, and nucleons  predicted by the
Z-burst mechanism for $m_\nu = 0.5$~eV, assuming sources
exclusively emitting neutrinos with fluxes equal for all
flavors. We show three cases of the source evolution parameter,
$m=-3, 0, 3$ by solid, dashed, and dotted lines, respectively.
Values assumed for the other parameters are:
$B=10^{-9}$ G, minimal URB strength, 
$z_{\rm min}=0$, $z_{\rm max}=3$, $E^\nu_{\rm max}=2 \times 10^{22}$ eV,
$q_\nu=1$. For each case the neutrino flux amplitude $f$
is obtained from minimizing $\chi^2$ for
$E_{\rm min} = 2.5 \times 10^{19}~{\rm eV}$.
Also shown are experimental upper limits on $\gamma-$ray
and neutrino fluxes (see text and Ref.~\cite{bs} for more
details).}
\label{F1}
\end{figure}

For a given set of values for all these parameters 
we find the neutrino flux amplitude $f$ in Eq.(\ref{para_nu}) 
obeying all experimental bounds on photon and neutrino fluxes
and explaining the UHECR flux at highest energies above some value
$E_{\rm min}$ by the secondary UHE protons and photons
by a maximum likelihood fit. The fit quality is characterized
by a $\chi^2$ value. Note that there are many different kinds
of extragalactic sources which can contribute
to the observed UHECR flux with energies below the GZK cutoff
$E_{\rm GZK}\simeq4 \times 10^{19}$ eV. Thus, one should take
$E_{\rm min} \lesssim E_{\rm GZK}$ if one wants to explain
all UHECR data above the cutoff by the Z-burst model.

Fig.~\ref{F1} illustrates how unrealistically high 
local neutrino background over-density could be avoided
by assuming sources that are more abundant at low redshifts.
In this figure we show primary neutrino and secondary  proton and
photon fluxes for the case $m_\nu = 0.5$~eV.
The following values have been assumed for the parameters of
Eq.~(\ref{para_nu}): spectral index $q_\nu = 1$, maximal
neutrino energy $E^\nu_{\rm max} = 2 \times 10^{22} ~{\rm eV}$,
minimal and maximal redshifts $z_{\rm min} = 0 $ and $z_{\rm max} = 3$. 
Three cases, corresponding to  the redshift
evolution index $m = -3, 0, 3$ are plotted as solid,
dashed and dotted lines, respectively. A typical value,
$B=10^{-9}\,$G, is assumed for the EGMF as well as the minimal
strength consistent with observations for the poorly known
URB~\cite{pb}. The latter results
in optimistic predictions for the UHE $\gamma-$ray flux.
The neutrino flux amplitude was fitted as described above
for $E_{\rm min} = 2.5 \times 10^{19}~{\rm eV}$.
Thus we require that the Z-burst model contributes to 9
bins of non-vanishing flux in the AGASA data.
For all three cases in Fig.~\ref{F1} we obtained $\chi^2(9)\simeq4$.

As one can see from Fig.~\ref{F1}, the value of the source evolution parameter
$m$ mainly affects photons with GeV energies. The value $m=3$
which was chosen in previous work~\cite{zburst2} is similar
to evolution of active galaxies.  The secondary photon
flux for such a source distribution is in conflict with
the diffuse GeV photon background observed by the EGRET experiment.
The uniform source distribution, $m = 0$, is already
in agreement with the EGRET flux, while negative values of the flux,
$m = -3$, lead to GeV photon fluxes well below it.
The latter case corresponds to sources which 
are more abundant now than at high redshifts. For example, BL Lacertae
objects for which correlations with UHECRs were found in
Ref.~\cite{corr_bllac}, are distributed in such a way.
Note that the choice of unknown EGMF strength $B$
and the URB flux affect only the UHE flux of photons and 
is not important in the EGRET region which is only
sensitive to the total injected EM energy.

\begin{figure}
\begin{center}
\leavevmode \epsfxsize=3.5in \epsfbox{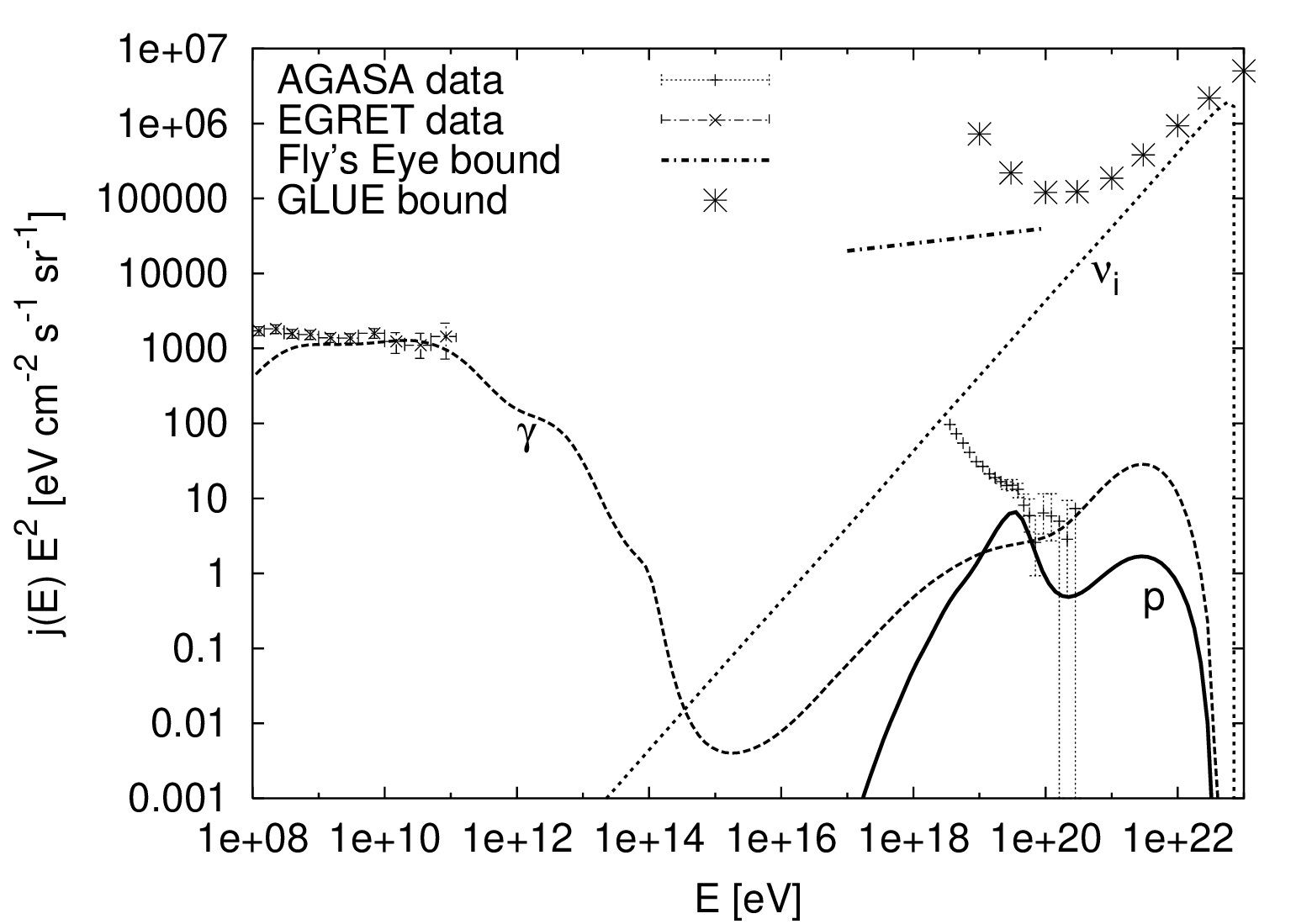}
\end{center}
\begin{center}
\leavevmode \epsfxsize=3.5in \epsfbox{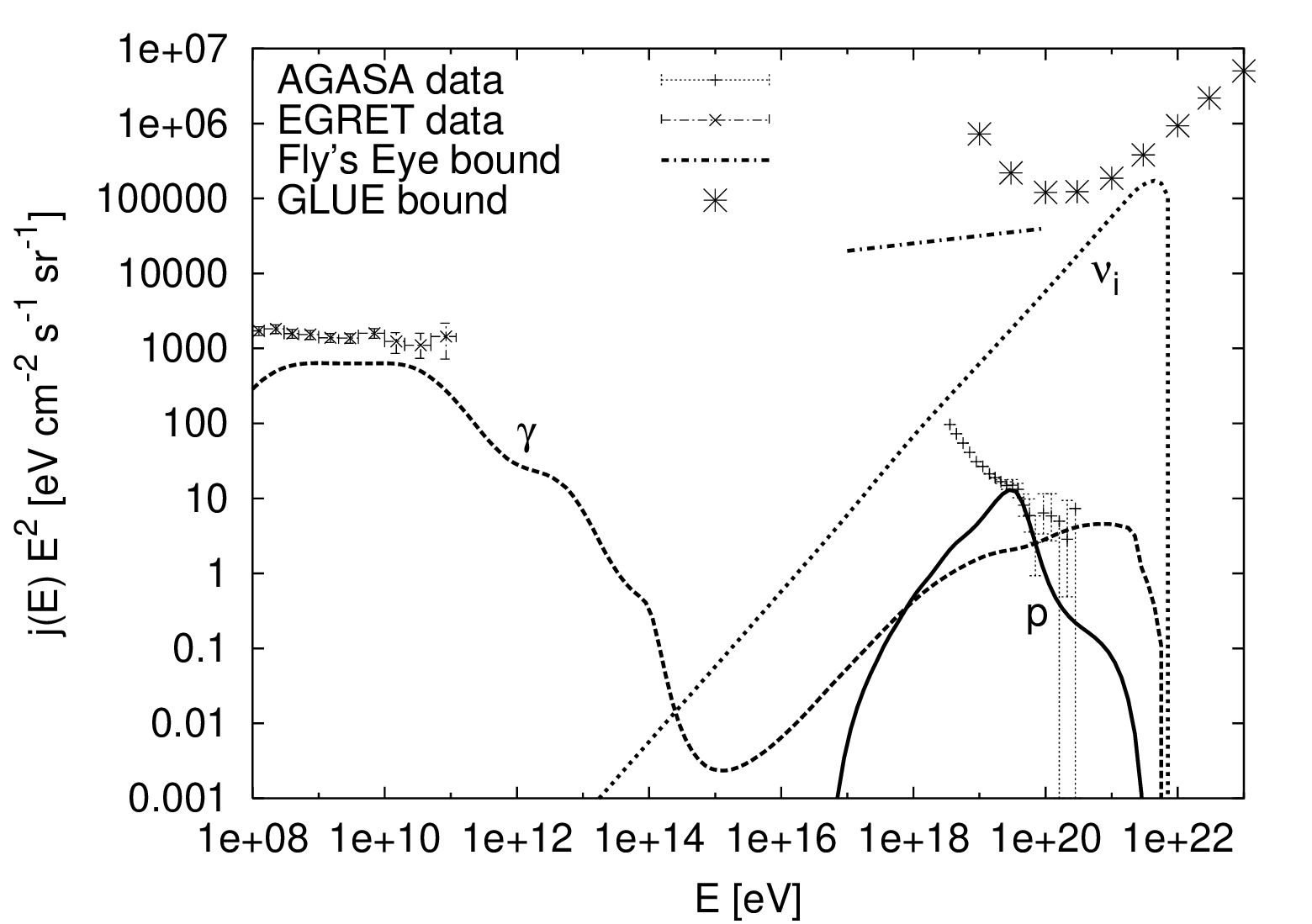}
\end{center}
\caption[...]{Dependence of flux predictions in the Z-burst model
on neutrino mass. (a) Small neutrino mass $m_\nu = 0.1$~eV,
with $m=-3$, $B=5 \times 10^{-11} G$, $E^\nu_{\rm max} = 10^{23} ~{\rm eV}$.
(b) Large neutrino mass $m_\nu = 1$~eV, with $m=0$, 
$B=10^{-12} G$, $E^\nu_{\rm max} = 10^{22} ~{\rm eV}$.
See also Fig.~\ref{F1} for explanations.}
\label{F2}
\end{figure}

Fig.~\ref{F2} shows results for varying neutrino masses with the
unknown parameters $B$, radio background strength, as well as
$f$, $q_\nu$, $E^\nu_{\rm max}$, $z_{\rm min}$, $z_{\rm max}$, and 
$m$ chosen such as to produce fits to the UHECR data of
comparable quality. We present in Fig.~\ref{F2} two extreme cases
of small  $m_\nu = 0.1$~eV and high $m_\nu = 1$~eV neutrino masses.
Because parameter space is huge, we fix some parameters to given values 
($z_{\rm min} = 0$, $z_{\rm max} = 3$, and minimal URB strength) 
and vary only the evolution parameter $m$, the EGMF strength $B$,
and the maximum energy $E^\nu_{\rm max}$ for every given neutrino mass.
Again, we determine the neutrino flux amplitude $f$ from
minimizing $\chi^2$. For $m_\nu = 0.1$~eV we get $\chi^2(6) = 1.6$, while for 
$m_\nu = 1$~eV we have $\chi^2(9) = 2.6$, i.e. the fit qualities
are comparable. For all intermediate masses we also find similar
fit qualities. From this we conclude that no preferred values for
the neutrino masses $m_\nu$ can be extracted. Instead, for every
neutrino mass exists a large parameter region in which the Z-burst
model with pure neutrino sources may work.

In Fig.~\ref{F2}a we show the case of small neutrino mass, $m_\nu = 0.1$~eV,
with $m=-3$, $B=5 \times 10^{-11} G$, and
$E^\nu_{\rm max} = 10^{23} ~{\rm eV}$. For small neutrino masses 
the resonance energy is large and thus secondary photons and protons
are produced at higher energies, apart from the photons produced
by t-channel leptons. Due to electro-magnetic cascades most of
the EM energy ends up in the GeV region and thus the EGRET
flux gives the most stringent bound. In particular, for $m_\nu< 0.1$~eV,
the parameter space for the Z-burst model shrinks, and
even a $m=-3$ distribution of sources is not consistent with the
data. However, there is no pronounced cutoff for photons
in this case (due to the $\alpha<1$ power law, see details in
Ref.~\cite{photons}). This allows to explain some fraction of the
UHECR events by photons. Finally, the required neutrino flux
is higher for small neutrino masses which makes the GLUE
experiment bound on the UHE neutrino flux~\cite{goldstone}
a crucial constraint for the Z-burst model.

In  Fig.~\ref{F2}b we present the case of  large neutrino mass,
$m_\nu = 1$~eV, with $m=0$, $B=10^{-12} G$, and
$E^\nu_{\rm max} = 10^{22} ~{\rm eV}$. In this case the required
flux of neutrinos is somewhat smaller. The available parameter
space for the Z-burst mechanism is large and the EGRET bound on
the GeV photon flux can be met even for a uniform distribution of sources.

\section{Photon and neutrino source}
\label{phot+nu}

It is well known that sources capable of accelerating
UHECRs produce $\gamma-$rays up to at least $E^\gamma_{\rm max}
\sim100\,$TeV~\cite{agn_obs}. Since in acceleration scenarios both
$\gamma-$rays and neutrinos
are produced as secondaries, the power $f_\gamma\sim1$ radiated
in $\gamma-$rays relative to neutrinos has to be of order unity.
For the $\gamma-$ray injection spectrum this implies
\begin{eqnarray}
  \phi_\gamma(E,z)&=&6ff_\gamma\,(1+z)^m\,
  \frac{g(q_\nu,E^\nu_{\rm max})}{g(q_\gamma,E^\gamma_{\rm max})}
  \,E^{-q_\gamma}\,\Theta(E^\gamma_{\rm max}-E)\nonumber\\
  &&z_{\rm min}\leq z\leq z_{\rm max}\,.\label{para_gamma}
\end{eqnarray}
Here, $q_\gamma$ is the $\gamma-$ray spectral index, and
$g(q,E_{\rm max})\equiv\int_{E_{\rm min}}^{E_{\rm max}}dE\,E^{1-q}$,
where we have introduced some small low energy cut-off $E_{\rm min}$
for convergence. 

\begin{figure}
\centering\leavevmode \epsfxsize=3.5in \epsfbox{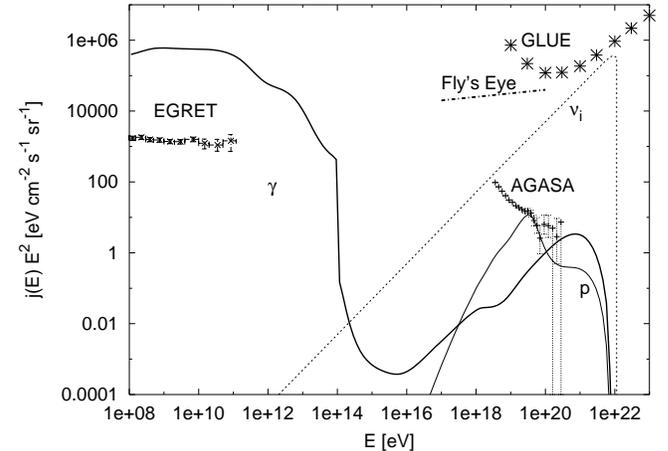}
\caption[...]{Same as Fig.~\ref{F1} 
($m_\nu=0.5$ eV)  for sources emitting equal power in neutrinos
and $\gamma-$rays, $f_\gamma=1$ in Eq.~(\ref{para_gamma}).
The other parameters are chosen as  $m=0$, $q_\gamma=2$,
$E^\gamma_{\rm max}=100\,$TeV.}
\label{F3}
\end{figure}

In this section we consider the more realistic case where the sources
also emit $\gamma-$rays with a power comparable to the
emitted neutrino power, $f_\gamma=1$, up to
$E^\gamma_{\rm max}=100\,$TeV in Eq.~(\ref{para_gamma}).
For all other parameters we chose values that minimize
the tension with the observational upper limits on the
UHE neutrino flux and the diffuse GeV $\gamma-$ray flux.
Fig.~\ref{F3} shows that in this case the Z-burst scenario
cannot be made consistent with observations. A possible 
solution to this problem is to down-scatter most of the
EM energy into the sub-MeV range within the source. Only
in this case can the EGRET bound be satisfied.
This would requires a very strong photon field up to $\gtrsim\,$keV
within the source.

Even the scenario in Fig.~\ref{F3} is still optimistic because it
assumes that the source is completely opaque to the
primary nucleons. While this may be
achieved easier than containment of $\gamma-$rays, for
example, by magnetic fields, it is clear from Fig.~\ref{F3}
that even if only a small fraction of the primary nucleons
leave the source, the nucleon flux between $\simeq10^{18}\,$eV
and $\simeq10^{19}\,$eV would be much higher than observed,
in agreement with the conclusions of Ref.~\cite{waxman}. This
problem could be avoided if the protons are deflected strongly enough
so that they could not reach the Earth. However, this
possibility also appears unrealistic as has been discussed
in Ref.~\cite{wb}. This problem can only be solved
if the protons are trapped in the source.

Finally we note that the Z-burst mechanism also poses extreme
requirements on the acceleration mechanism itself since the primary
protons have to be accelerated to energies
$E^p_{\rm max}\sim10E^\nu_{\rm max}\gtrsim4\times10^{22}
\,{\rm eV}\,({\rm eV}/m_\nu)$. In contrast, known mechanisms
are usually limited to $E^p_{\rm max}\lesssim10^{22}\,$eV~\cite{bier-rev}.

Thus, the Z-burst model imposes the following requirements
onto the sources: They should emit energy only in neutrinos
and in sub-MeV $\gamma-$rays, and should also trap most of the
primary protons.

\section{Conclusions}
The Z-burst mechanism where the highest energy cosmic rays
are produced by neutrino beams interacting with the relic
neutrino background only works with sources exclusively
emitting neutrinos in the ultra-high energy regime. In order
to avoid conflict with the known diffuse backgrounds of $\gamma-$rays,
these sources should emit photons only in the sub-MeV region.
In addition, they should trap primary protons in order to avoid
an excessive nucleon flux from the source, and should be able to
accelerate these protons up to $E^p_{\rm max} \gtrsim 10^{23}
\,{\rm eV}\,({\rm eV}/m_\nu)$. None of the astrophysical
acceleration models existing in the literature seems to
meet this requirement.

Under the assumption that such an extreme source class
nevertheless exists we have shown that the Z-burst mechanism
can work without unrealistically high local relic neutrino over-densities
if the neutrino sources are typically more abundant at present than in
the past. Especially neutrino masses $m_\nu\lesssim0.5\,$eV
require a non-uniform source distribution $\propto(1+z)^m$
with negative evolution factor, $m<0$, as is the case with
BL Lacertae objects.

The contribution to the UHECR flux from such a speculative extragalactic
neutrino source class due to the Z-burst mechanism would exhibit a GZK-cutoff
for nucleons and would be dominated by $\gamma-$rays at higher energies.
Furthermore, the required UHE neutrino fluxes are close to existing
upper limits and should be easily
detectable by future experiments such as Auger~\cite{auger},
Euso~\cite{euso}, RICE~\cite{rice}, or by other radio detection techniques~\cite{radhep}.

The space of parameters characterizing neutrino sources and their evolution
is highly degenerate when fluxes are fit to the observed
UHECR fluxes. Since evidence of relic neutrinos and extraction of
absolute neutrino masses requires conservative assumptions about these
unknown parameters, we conclude that the current state
of knowledge does not allow to extract any meaningful
information on neutrino masses from UHECR data.

\section*{Acknowledgements}
We would like to thank Zoltan Fodor, Sandor Katz, Andreas Ringwald and
Tom Weiler for detailed discussions on this subject.
We are grateful to Otmar Biebel for providing us with Z-decay data
and to Joel Primack and James Bullock for making available to us their
results for the infrared/optical background in electronic form.


\end{multicols}

\end{document}